\RequirePackage{lineno}
\documentclass[a4paper,10pt,oneside]{cpc-hepnp}
\usepackage{multicol}
\usepackage{graphicx}
\usepackage{booktabs}
\usepackage{bm}
\usepackage{mathrsfs}
\usepackage{bbm}
\usepackage{amscd}
\usepackage[tbtags]{amsmath}
\usepackage{amssymb,bm,mathrsfs,bbm,amscd}
\usepackage{lastpage}
\usepackage{upgreek,fancyhdr}
\usepackage{color}
\usepackage{geometry}
\usepackage{xspace}
\begin{document}
\newcommand{\calL}{\ensuremath{\mathcal{L}}}
\newcommand{\jpsi}{\ensuremath{\rm J/ \uppsi}\xspace}

\fancyhead[co]{\footnotesize BESIII Col.:
{\boldmath Number of \jpsi events at BESIII}} \footnotetext[0]{Received }
\title{\boldmath Number of \jpsi events at BESIII\thanks{Supported
in part by National Key R\&D Program of China under Contracts
Nos. 2020YFA0406300, 2020YFA0406400; National Natural Science
Foundation of China (NSFC) under Contracts Nos. 11625523, 11635010,
11735014, 11822506, 11835012, 11935015, 11935016, 11935018,
11961141012, 12022510, 12025502, 12035009, 12035013,
12061131003,12075252; the Chinese Academy of Sciences (CAS)
Large-Scale Scientific Facility Program; Joint Large-Scale Scientific
Facility Funds of the NSFC and CAS under Contracts Nos. U1732263,
U1832207; CAS Key Research Program of Frontier Sciences under Contract
No. QYZDJ-SSW-SLH040; 100 Talents Program of CAS; INPAC and Shanghai
Key Laboratory for Particle Physics and Cosmology; ERC under Contract
No. 758462; European Union Horizon 2020 research and innovation
programme under Contract No. Marie Sklodowska-Curie grant agreement No
894790; German Research Foundation DFG under Contracts Nos. 443159800,
Collaborative Research Center CRC 1044, FOR 2359, GRK 214; Istituto Nazionale di Fisica Nucleare, Italy; Ministry of Development of Turkey under Contract No. DPT2006K-120470; National Science and Technology fund; Olle Engkvist Foundation under Contract No. 200-0605; STFC (United Kingdom); The Knut and Alice Wallenberg Foundation (Sweden) under Contract No. 2016.0157; The Royal Society, UK under Contracts Nos. DH140054, DH160214; The Swedish Research Council; U. S. Department of Energy under Contracts Nos. DE-FG02-05ER41374, DE-SC-0012069.
}
\vspace{-0.5in}
}
\maketitle

\begin{small}
\begin{center}
\begin{small}
\begin{center}
M.~Ablikim$^{1}$, M.~N.~Achasov$^{10,b}$, P.~Adlarson$^{68}$, S. ~Ahmed$^{14}$, M.~Albrecht$^{4}$, R.~Aliberti$^{28}$, A.~Amoroso$^{67A,67C}$, M.~R.~An$^{32}$, Q.~An$^{64,50}$, X.~H.~Bai$^{58}$
% ...
\\
\vspace{0.2cm}
(BESIII Collaboration)\\
\vspace{0.2cm} {\it
$^{1}$ Institute of High Energy Physics, Beijing 100049, People's Republic of China\\
% ...
$^{4}$ Bochum Ruhr-University, D-44780 Bochum, Germany\\
% ...
$^{10}$ G.I. Budker Institute of Nuclear Physics SB RAS (BINP), Novosibirsk 630090, Russia\\
% ...
$^{14}$ Helmholtz Institute Mainz, Staudinger Weg 18, D-55099 Mainz, Germany\\
% ...
$^{28}$ Johannes Gutenberg University of Mainz, Johann-Joachim-Becher-Weg 45, D-55099 Mainz, Germany\\
% ...
$^{32}$ Liaoning Normal University, Dalian 116029, People's Republic of China\\
% ...
$^{50}$ State Key Laboratory of Particle Detection and Electronics, Beijing 100049, Hefei 230026, People's Republic of China\\
% ...
$^{58}$ University of Jinan, Jinan 250022, People's Republic of China\\
% ...
$^{64}$ University of Science and Technology of China, Hefei 230026, People's Republic of China\\
% ...
$^{67}$ University of Turin and INFN, (A)University of Turin, I-10125, Turin, Italy; (B)University of Eastern Piedmont, I-15121, Alessandria, Italy; (C)INFN, I-10125, Turin, Italy\\
$^{68}$ Uppsala University, Box 516, SE-75120 Uppsala, Sweden\\
% ...
$^{b}$ Also at the Novosibirsk State University, Novosibirsk, 630090, Russia\\
% ...
}\end{center}

\vspace{0.4cm}
\end{small}

\end{center}
\vspace{-0.2cm}
\end{small}
\begin{abstract} Using
inclusive decays of the \jpsi, a precise determination of the
number of \jpsi events collected with the BESIII detector is
performed.  For the two data sets taken in 2009 and 2012, the numbers
of \jpsi events are recalculated to be $(224.0 \pm
1.3)\times10^6$ and $(1088.5 \pm 4.4)\times10^6$ respectively, which
are in good agreement with the previous measurements.  For the
\jpsi sample taken in 2017--2019, the number of events is
determined to be $(8774.0 \pm 39.4)\times10^{6}$.  The total number
of \jpsi events collected with the BESIII detector is
determined to be $(10087 \pm 44)\times10^{6}$, where the uncertainty
is dominated by systematic effects and the statistical uncertainty is
negligible.  \end{abstract}


\begin{multicols}{2}
\section{Introduction}

As a charmonium ground state, the \jpsi offers a unique
laboratory for studying light hadron spectroscopy. In particular,
\jpsi decays can be used to search for exotic hadrons composed of
light quarks and gluons, which are key to a fuller understanding
of the nature of the strong interaction.

Many important results in light hadron spectroscopy~\cite{white} have
been reported based on $(1310.6 \pm 7.0)\times 10^6$ \jpsi
events collected by the BESIII experiment~\cite{bes3} in 2009 and
2012. An additional large sample of \jpsi events was
collected by BESIII during 2017--2019 to improve the precision of the
measurements and search for new processes.  The three data samples of
\jpsi events collected at BESIII are summarized in
Table~\ref{samples}.

\begin{center}
\tabcaption{\label{samples}Data samples used in the determination of the
number of \jpsi events}
\footnotesize
\begin{tabular*}{80mm}[htbp]{l@{\extracolsep{\fill}}cccc}
\toprule Data set & $\sqrt{s}$ &$\calL_\text{online}$ & Date (duration) \\
 &&&(YYYY-MM-DD)\\\hline
    \jpsi  & 3.097 GeV& $2678$pb$^{-1}$ & 2017-08-12 -- 2019-06-02 \\
    QED1  & 3.08 GeV& $48$pb$^{-1}$ & 2018-04-12 -- 2018-04-14 \\
    QED2  & 3.08 GeV& $88$pb$^{-1}$ & 2019-02-07 -- 2019-02-11 \\
    $\rm \uppsi(3686)$  & 3.686 GeV& $25$pb$^{-1}$ & 2018-05-20 \\\hline
    \jpsi  & 3.097 GeV& $323$pb$^{-1}$ & 2012-04-10 -- 2012-05-22 \\
    QED1  & 3.08 GeV& $13$pb$^{-1}$ & 2012-04-08 \\
    QED2  & 3.08 GeV& $17$pb$^{-1}$ & 2012-05-23 -- 2012-05-24 \\
    $\rm \uppsi(3686)$  & 3.686 GeV& $7.5$pb$^{-1}$ & 2012-05-26 \\\hline
    \jpsi  & 3.097 GeV& $82$pb$^{-1}$ & 2009-06-12 -- 2009-07-28 \\
    QED  & 3.08 GeV& $0.3$pb$^{-1}$ & 2009-06-19 \\
    $\rm \uppsi(3686)$  & 3.686 GeV& $150$pb$^{-1}$ & 2009-03-07 -- 2009-04-14 \\\hline
\bottomrule
\end{tabular*}
\end{center}

This paper reports a precise determination of the total number of \jpsi
events, which is an important quantity for many analyses using these
data samples.
The number of \jpsi events for the new
samples collected in 2017-2019 is determined with the same method as
the one used in the previous measurements~\cite{njsi2012}. In
addition, in this analysis we also recalculate the number of
\jpsi events for the two data samples taken in 2009 and 2012,
reconstructed using
the latest BESIII software. The number of \jpsi events,
$N_{\jpsi}$, is calculated as
\begin{eqnarray} \label{Nojpsi}
 N_{\jpsi}=\frac{N_\text{sel}-N_\text{bg}}{\epsilon_\text{trig}
 \times \epsilon^{\uppsi(3686)}_\text{data} \times f_\text{cor}},
\end{eqnarray}
where $N_\text{sel}$ is the number of inclusive \jpsi decays selected
from the \jpsi data; $N_\text{bg}$ is the number of background events
estimated with continuum data taken at $\sqrt{s} = 3.08\,\text{GeV}$;
$\epsilon_\text{trig}$ is the trigger efficiency;
$\epsilon^{\uppsi(3686)}_\text{data}$ is the inclusive \jpsi detection
efficiency determined experimentally using the \jpsi sample from the
reaction $\rm \uppsi(3686) \rightarrow \uppi^+ \uppi^- \jpsi$.
$f_\text{cor}$ is a correction factor that accounts for the difference
in the detection efficiency between the \jpsi events produced at rest
and those produced in
$\rm \uppsi(3686)\rightarrow \uppi^+ \uppi^- J/\uppsi$. $f_\text{cor}$
is expected to be approximately unity, and is determined by the Monte
Carlo (MC) simulation sample with
\begin{eqnarray} \label{Fcor}
  f_\text{cor} = \frac {\epsilon^{\rm J/\uppsi}_\text{MC}}
  {\epsilon^{\uppsi(3686)}_\text{MC}}, \end{eqnarray}
where
$\epsilon^{\rm J/\uppsi}_\text{MC}$ is the detection efficiency of
inclusive \jpsi events determined from the MC sample of $\rm J/\uppsi$
events produced directly in electron-positron collisions, and
$\epsilon^{\uppsi(3686)}_\text{MC}$ is that from the MC sample of
$\rm \uppsi(3686)\to \uppi^+\uppi^- J/\uppsi$. For the number of
$\rm J/\uppsi$ events determined with Eq.~\ref{Nojpsi}, only
$f_\text{cor}$ depends on MC simulation. According to Eq.~\ref{Fcor},
the uncertainties related to MC simulation (including generator,
detector response etc.) almost cancel since they impact both the
numerator and denominator, which improves the precision of the number
of \jpsi events.  In the MC simulation, the production of \jpsi and
$\rm \uppsi(3686)$ resonances is simulated with a
GEANT4-based~\cite{geant4} MC software, which includes the geometric
description of the BESIII detector and the detector response. The
simulation models the beam energy spread and initial state radiation
(ISR) in the $e^+e^-$ annihilations with the generator
KKMC~\cite{kkmc}.  The known decay modes are modeled with
EVTGEN~\cite{evtgen,djl} using branching fractions taken from the
Particle Data Group~\cite{PDG}, and the remaining unknown charmonium
decays are modeled with LUNDCHARM~\cite{LUND,LUND2}.

\section{\boldmath Inclusive \jpsi selection criteria}
\label{hadsel}

Candidate events must contain two or more charged tracks which are
required to have a momentum less than 2.0 GeV/c and to be within a
polar angle ($\theta$) range of $|\cos \theta| < 0.93$, where
$\theta$ is defined with respect to the axis of the Main Drift Chamber (MDC).
The distance of closest approach to the
interaction point (IP) must be less than 15 cm along the $z$-axis,
$|V_z|$, and less than 1 cm in the transverse plane, $V_r$. Photon
candidates are identified using isolated showers in the electromagnetic
calorimeter (EMC). The deposited energy of each shower must be more
than 25 MeV in the barrel region ($|\cos\theta| < 0.83$) and more than
50 MeV in the end cap region ($0.86< |\cos\theta| < 0.93$). To
suppress electronic noise and showers unrelated to the event, the
difference between the EMC time and the event start time is required
to be within [0, 700] ns.

To suppress events from Quantum Electrodynamics (QED) processes
(\emph{i.e.} Bhabha and dimuon events), from cosmic rays, beam-induced
backgrounds and electronic noise, a series of selection criteria are applied to
the candidate events.

The sum of charged particle energies computed from the track momenta
assuming a pion mass and the neutral shower energies deposited
in the EMC, $E_\text{vis}$, is required to be greater than 1.0 GeV.
Figure~\ref{evis} shows a comparison of the $E_\text{vis}$
distribution between the \jpsi data, the data taken at
$\sqrt{s} = 3.08$ GeV, and the inclusive \jpsi MC sample. The
requirement $E_\text{vis} > 1.0~\text{GeV}$ removes one third of the
background events while retaining $99.5\%$ of the signal events.

\begin{center}
\includegraphics[width=8.cm,height=5.7cm]{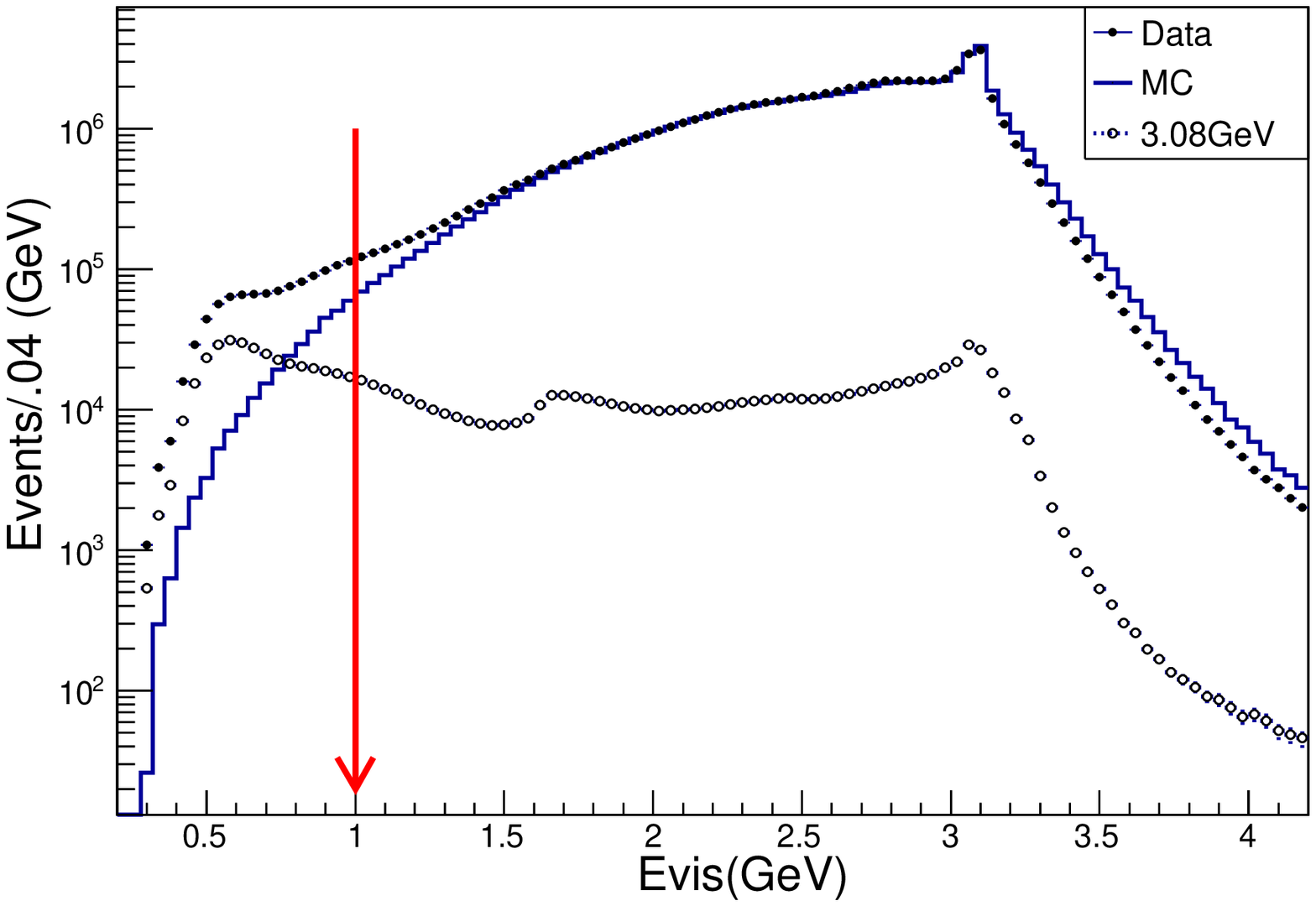}
\figcaption{\label{evis} Distributions of the visible energy $E_\text{vis}$ for \jpsi data (dots with error bars), continuum data
at $\sqrt{s}=3.08$ GeV (open circles with error bars, normalized to the integrated luminosity of \jpsi data) and MC simulation of inclusive \jpsi events (histogram). The arrow indicates the minimum $E_\text{vis}$ required to select inclusive events.  }
\end{center}

For events with only two charged tracks, the momentum of each track is
required to be less than 1.5 GeV/$c$ to exclude Bhabha and dimuon
events. Figure~\ref{pcutn} shows a scatter plot of the momenta of
the two charged tracks, and the solid lines depict the momentum
requirement. Figure~\ref{etrk} displays the distribution of energy
deposited by the charged particles in the EMC; a significant peak
around 1.5 GeV is from Bhabha events, which can be rejected by
requiring the energy deposited in the EMC be less than 1 GeV for each
charged track.
\begin{center}
\includegraphics[width=8.0cm,height=5.7cm]{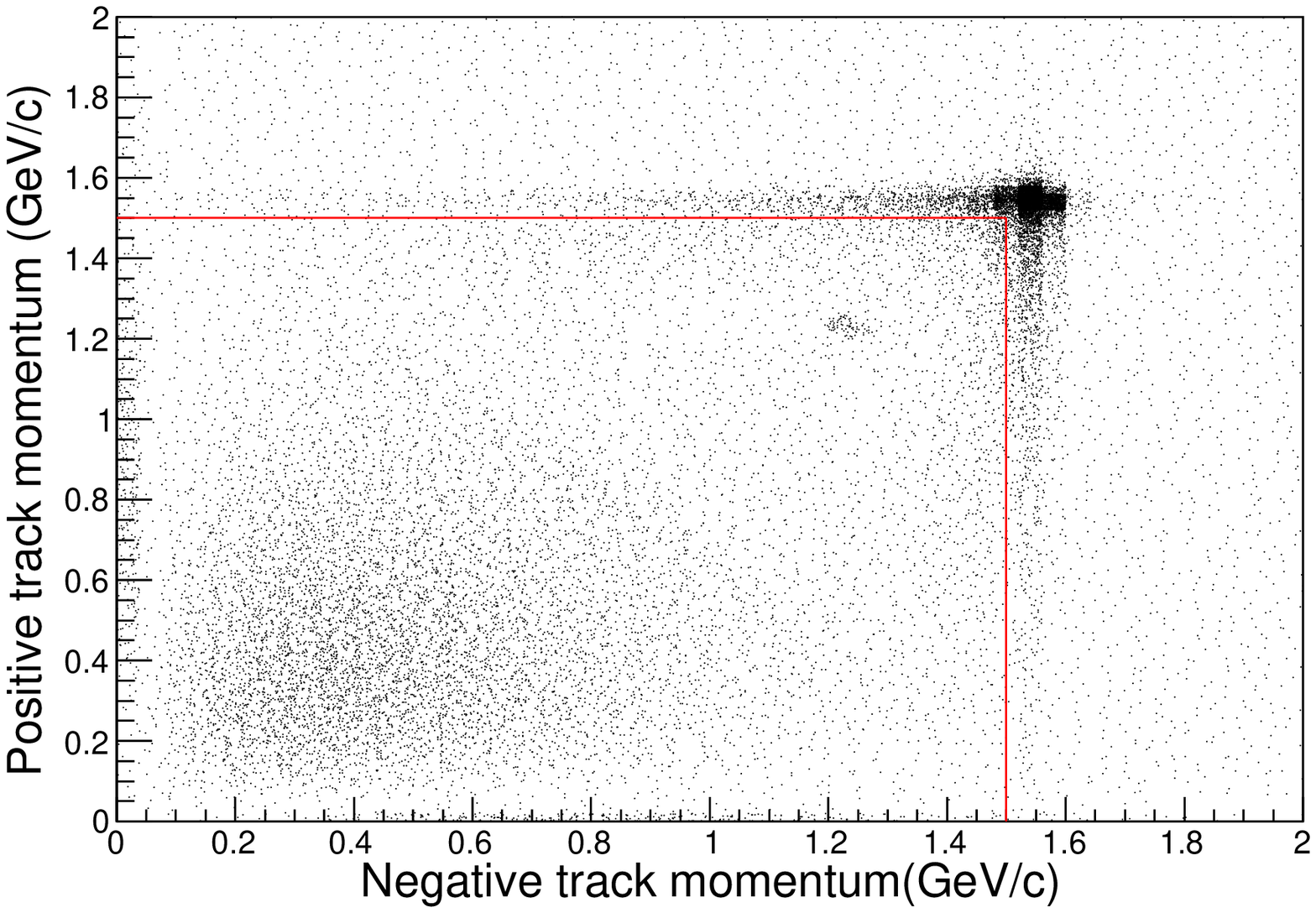}
\figcaption{\label{pcutn} Distribution of the momenta of the charged
  tracks for 2-prong events in data. The cluster around 1.55 GeV/$c$
  corresponds to the contribution from lepton pairs and the cluster at
  1.23 GeV/$c$ comes from $\rm J/\uppsi \rightarrow p\bar{p}$.  Most
  of lepton pairs are removed with the requirements on the two charged
  tracks, $p_1<1.5$ GeV/$c$ and $p_2<1.5$ GeV/$c$, as indicated by the
  solid lines.}
\end{center}
\begin{center}
\includegraphics[width=8.0cm,height=5.7cm]{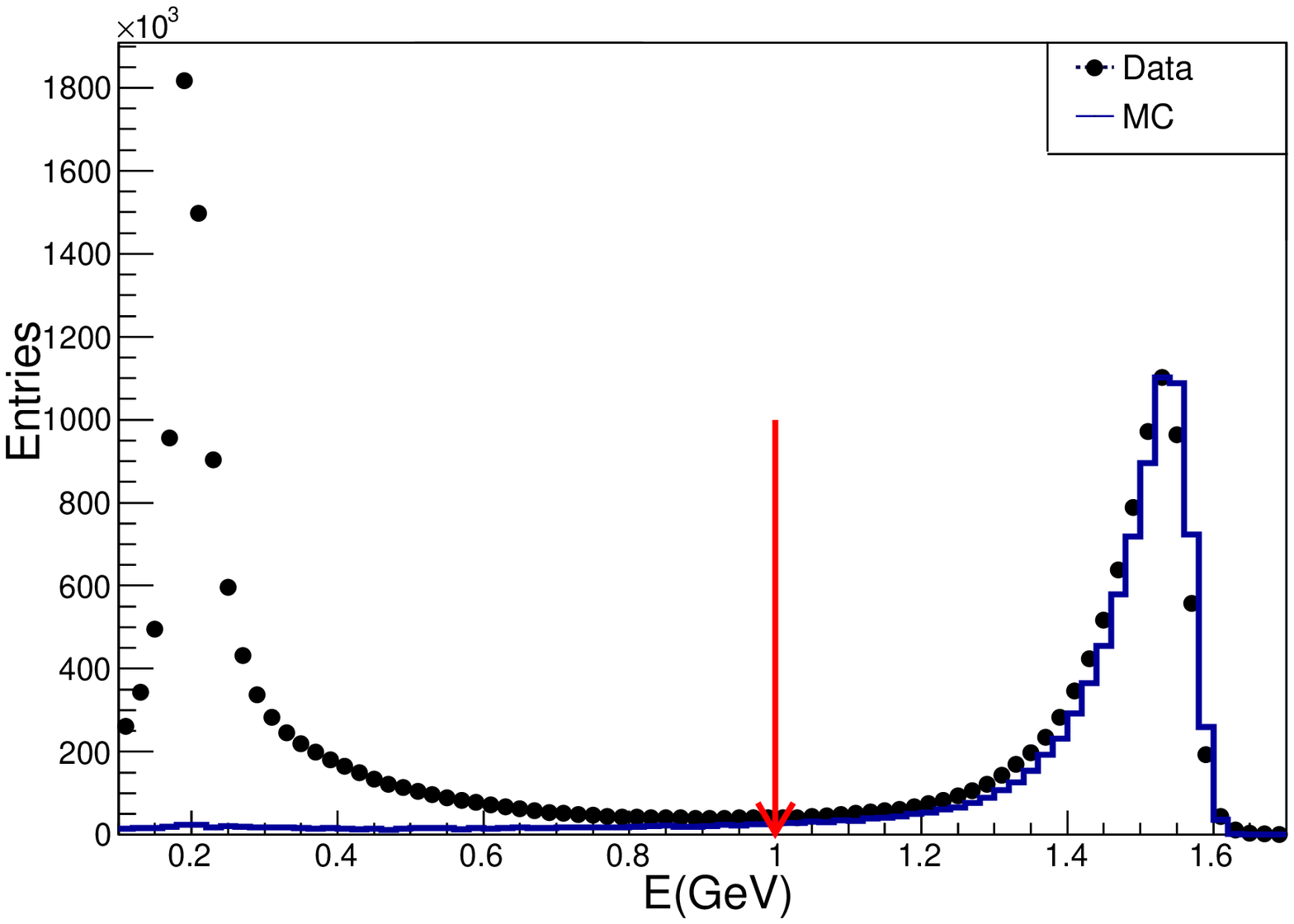}
\figcaption{\label{etrk} Distributions of deposited energy in the EMC
  for the charged tracks of 2-prong events for \jpsi data
  (dots with error bars) and for the combined, normalized MC
  simulations of $\rm e^+e^-\rightarrow e^+e^-(\gamma)$ and $\rm
  J/\uppsi\rightarrow e^+e^-(\gamma)$ (histogram).}
\end{center}

After the above requirements, $N_\text{sel} = (6912.03 \pm 0.08)\times
10^{6}$ candidate events are selected from the \jpsi data
taken in 2017--2019. The distributions of the track parameters of
closest approach along the beam line and in radial direction ($V_z$
and $V_r$), the polar angle ($\cos\theta$), and the total energy
deposited in the EMC ($E_\text{EMC}$) after subtracting background
events estimated with the continuum data taken at $\sqrt{s} = 3.08$
GeV (see Sec.~{\ref{bkg}} for details) are illustrated in
Fig.~\ref{vz0}. The multiplicity of charged tracks ($N_\text{good}$)
is shown in Fig.~\ref{ngood}, where the MC sample generated according
to the standard MC model agrees very well with the data while the MC
sample generated with an `incomplete' MC model deviates from the data. The
standard MC model includes the known decay processes listed in the PDG and
the unknown ones modeled with LUNDCHARM, while the incomplete MC model only
includes the known decay modes listed in the PDG.  The effect of this
discrepancy on the determination of the number of \jpsi
events is small, as described in Sec.~{\ref{syst}}.
\begin{center}
\includegraphics[width=8.0cm,height=5.0cm]{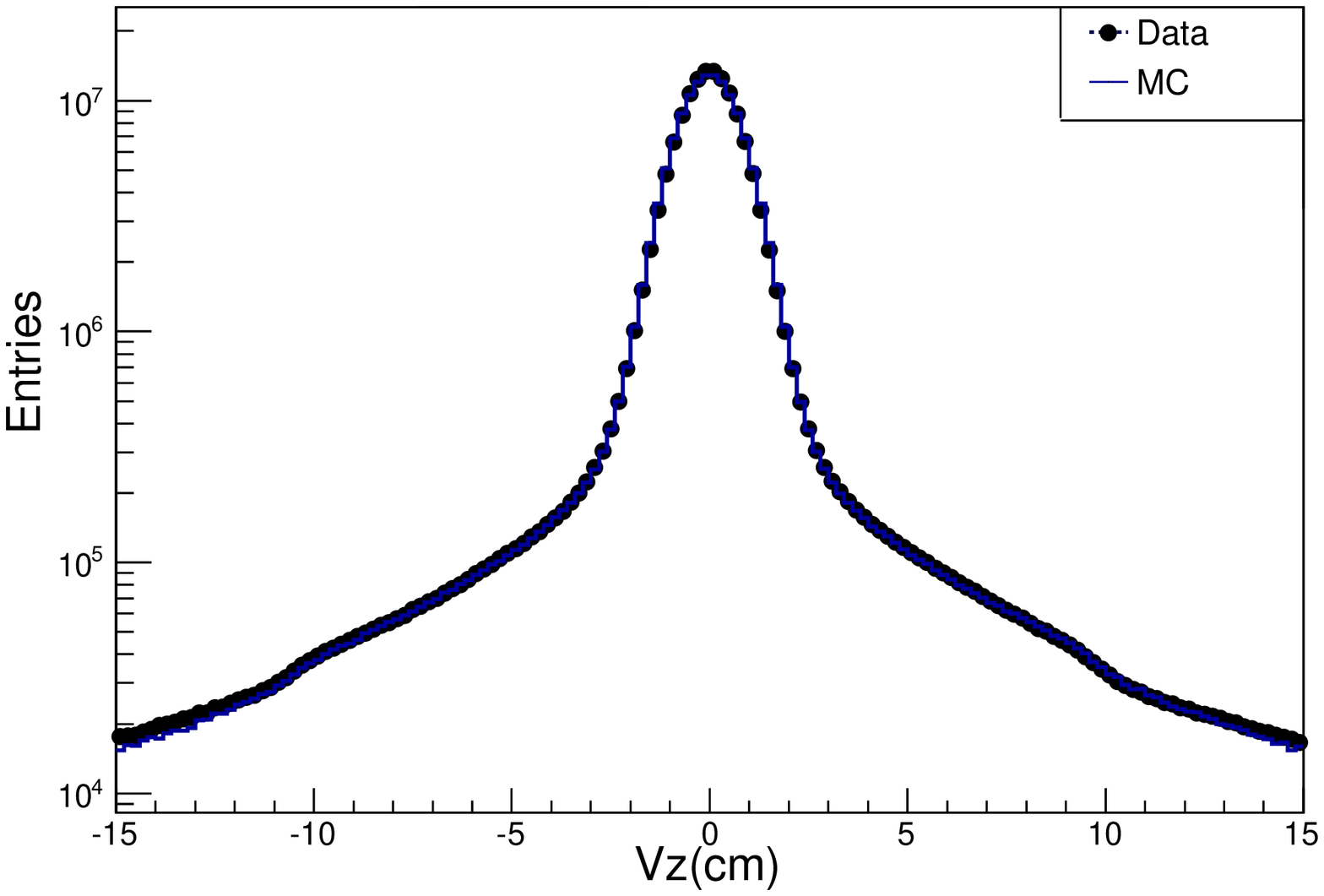}
\put(-175,110){(a)}\\
\includegraphics[width=8.0cm,height=5.0cm]{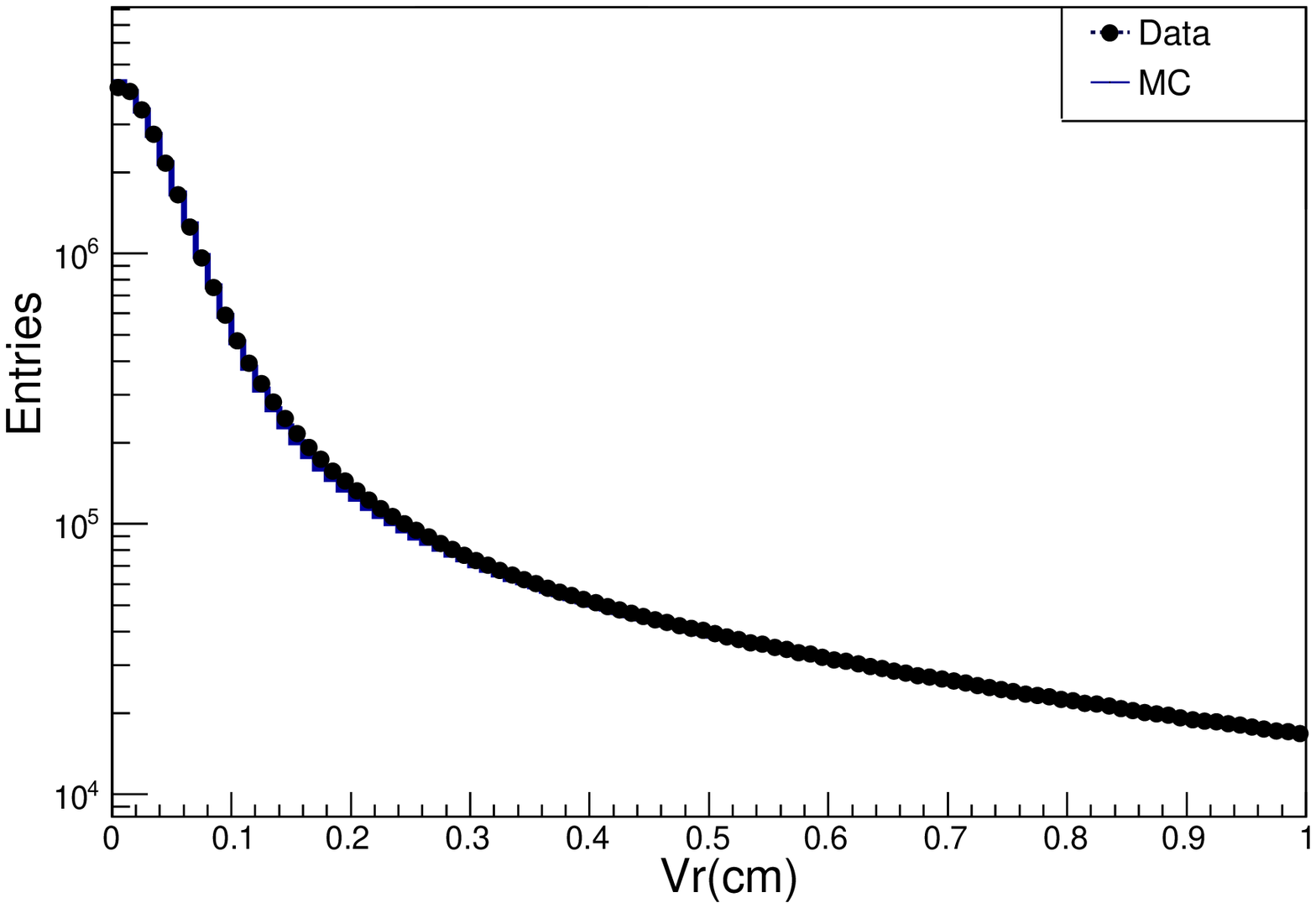}
\put(-175,110){(b)}\\
\includegraphics[width=8.0cm,height=5.0cm]{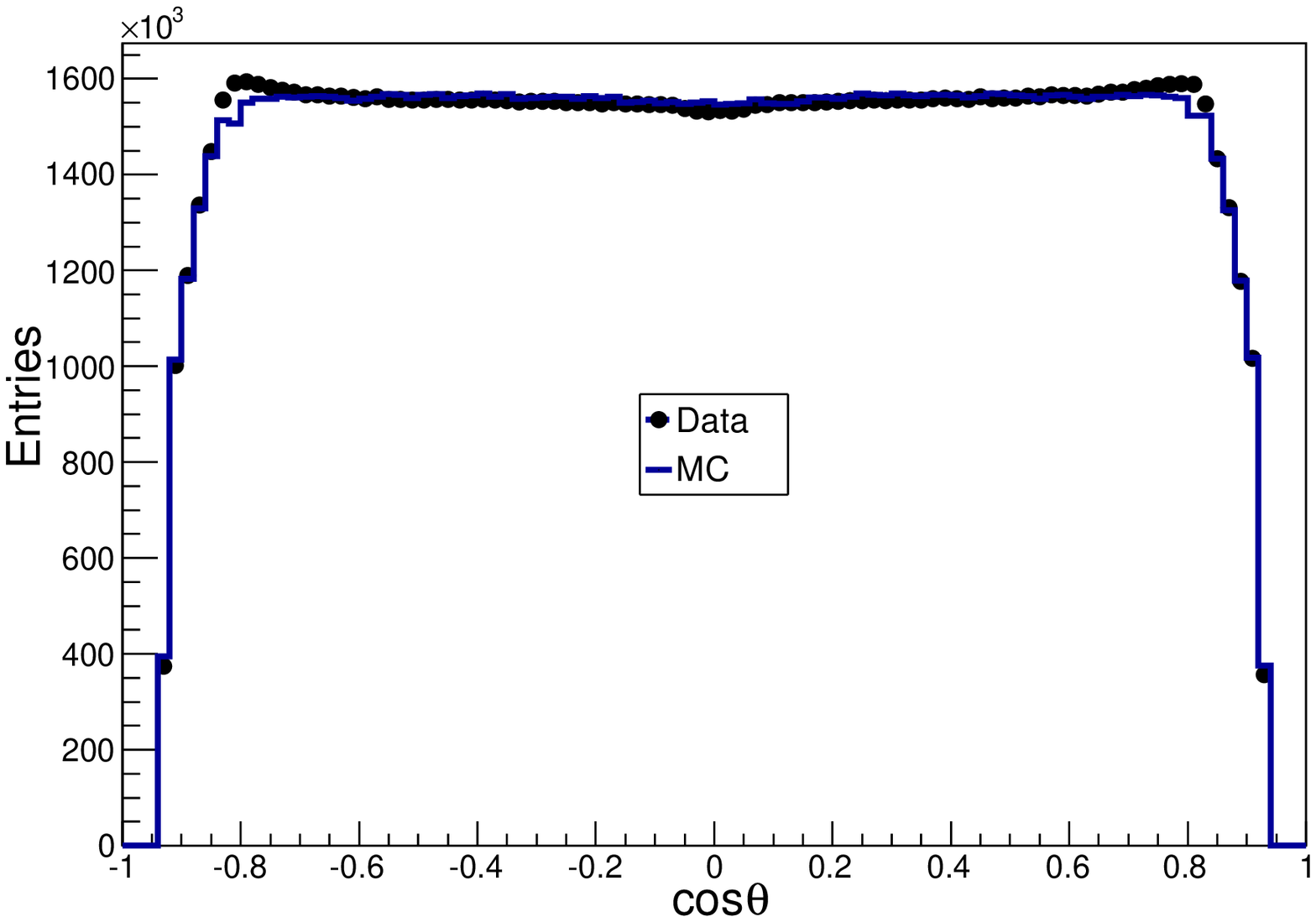}
\put(-175,110){(c)}\\
\includegraphics[width=8.0cm,height=5.0cm]{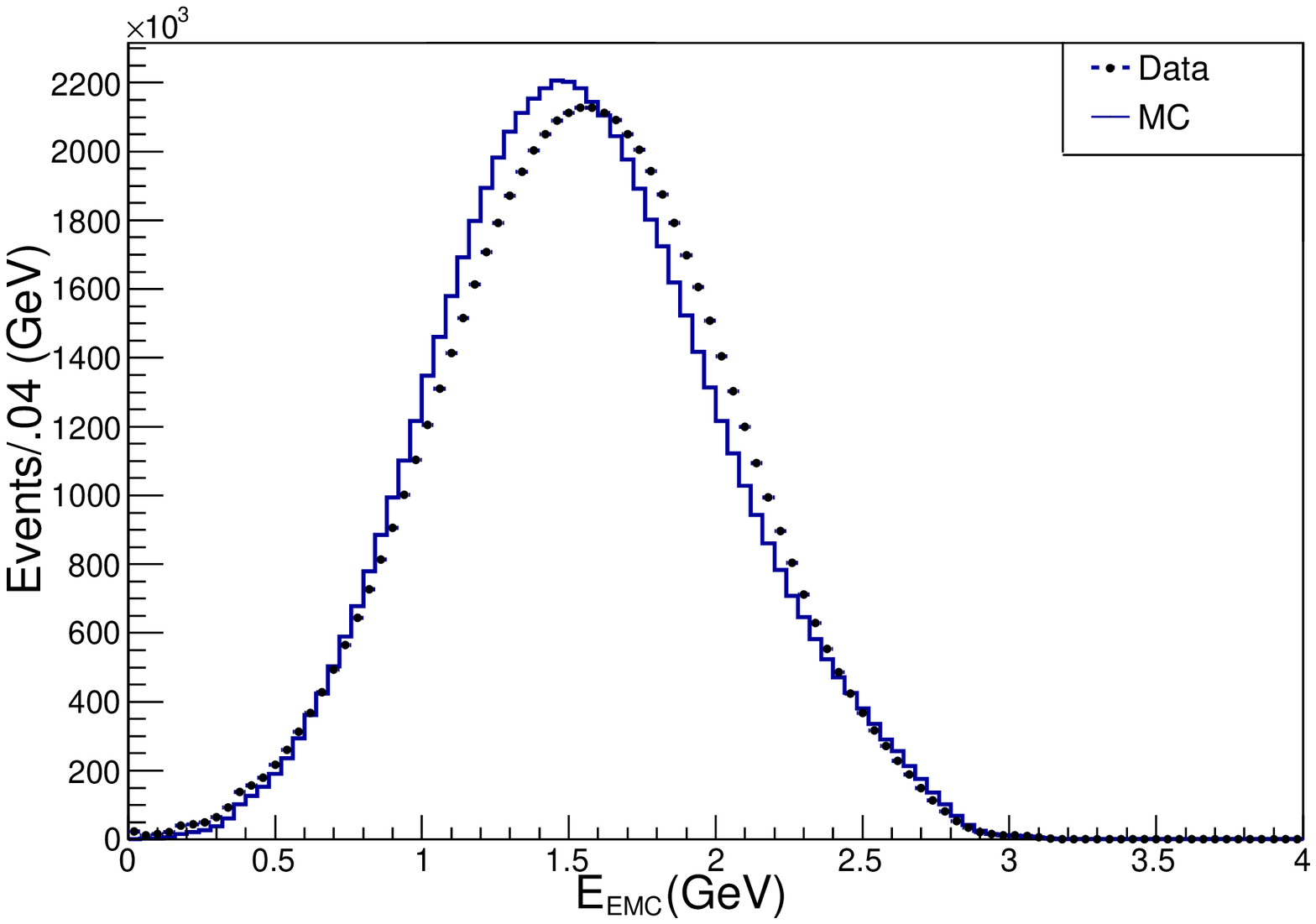}
\put(-175,110){(d)}
\figcaption{\label{vz0} Comparison of distributions between \jpsi data (dots with error bars) and MC simulation of inclusive \jpsi (histogram): (a) $V_z$, (b) $V_r$, (c) $\cos\theta$ of charged tracks, (d) total energy deposited in the EMC.}
\end{center}

\begin{center}
\includegraphics[width=8.0cm,height=5.7cm]{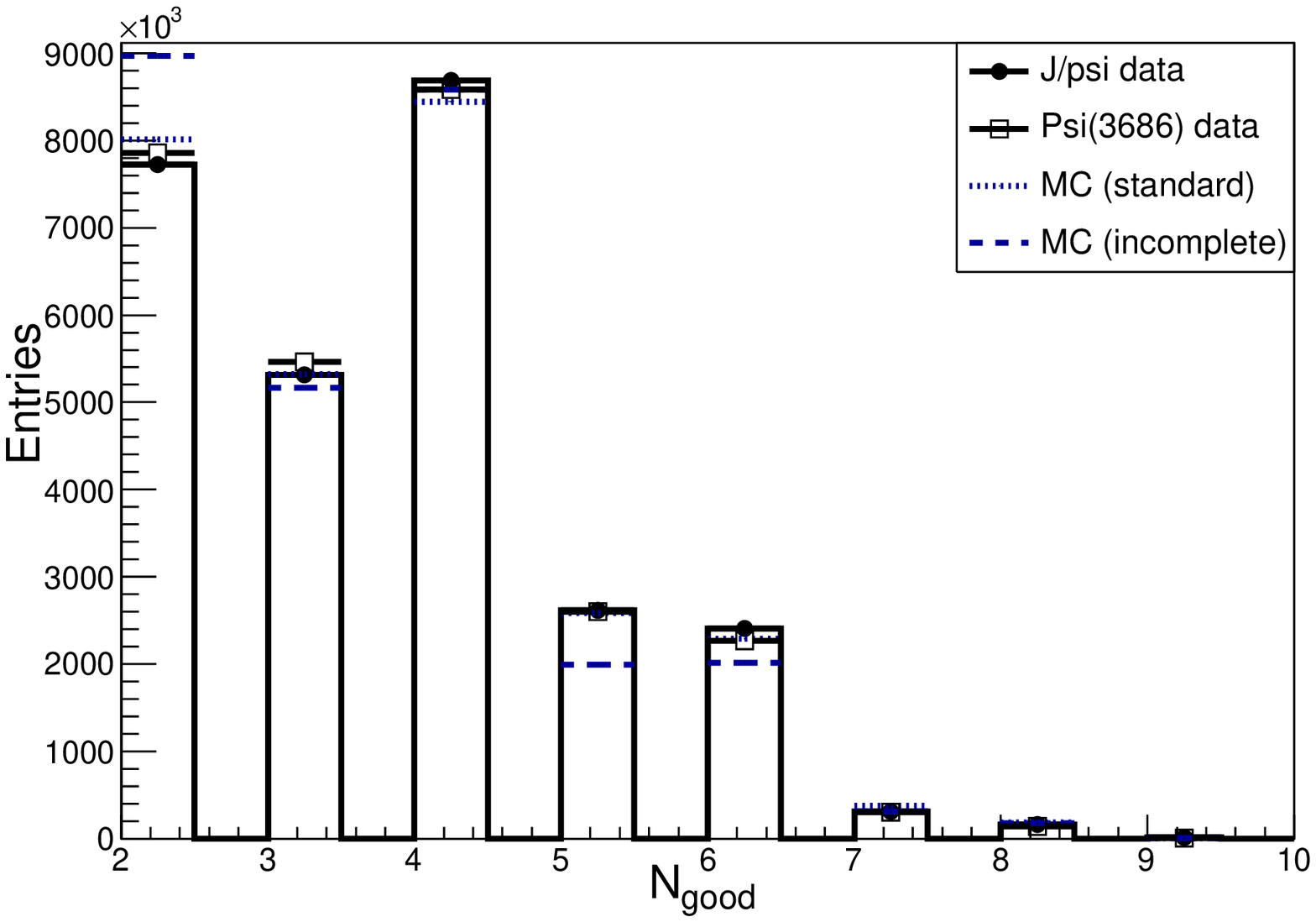}
\figcaption{\label{ngood} Distributions of the reconstructed charged
  track multiplicity in inclusive \jpsi events for \jpsi
  data (dots with error bars) and $\rm \uppsi(3686)$ data
  (squares with error bars) and MC simulation generated using standard
  and incomplete MC model (short-dashed and long-dashed histograms,
  respectively). }
\end{center}

\section{Background analysis}
\label{bkg}

In this analysis, the potential background sources include QED
processes, beam-induced background, cosmic rays, and electronic noise. The
continuum data samples at $\sqrt{s}=3.08$ GeV are taken in
close chronological order to each \jpsi sample to estimate
these backgrounds.

The integrated luminosity is determined using the process
$\rm e^+e^-\rightarrow \gamma\gamma$. The candidate events are
selected by requiring at least two showers in the EMC within
$|\cos\theta | < 0.8$ and with the energy of the second most energetic shower
between 1.2 and 1.6 GeV.  The number of signal events is determined
from the number of events in the signal region
$|\Delta \phi| < 2.5^\circ$, and the background is estimated from
those in the side-band region $2.5^\circ < |\Delta \phi| < 5^\circ$,
where $\Delta \phi = |\phi_{\gamma1} - \phi_{\gamma2}| - 180^\circ $
and $\phi_{\gamma1/2}$ are the azimuthal angles of the two photon candidates. Taking into
account the detector efficiency obtained from the MC simulation and
the cross section of the QED process
$\rm e^+e^- \rightarrow \gamma\gamma$, the integrated luminosities of
the \jpsi data sample and the sample taken at $\sqrt{s}=3.08$~GeV in
2017--2019 are determined to be $(2568.07 \pm 0.40) \;\text{pb}^{-1}$
and $(136.22 \pm 0.09)\;\text{pb}^{-1}$, respectively, where the
errors are statistical only.

After applying the same selection criteria as for the \jpsi
data, $N_{3.08}=6,363,941 \pm 2,523$ events are selected from the
continuum data taken at $\sqrt{s} = 3.08$ GeV. Assuming the same
detection efficiency at $\sqrt{s} = 3.08$ GeV as for the \jpsi
peak and taking into account the energy-dependent cross
section of the QED processes, the number of background events for the
\jpsi sample, $N_\text{bg}$, is estimated to be
\begin{eqnarray}
\label{Nbg} N_\text{bg}=N_{3.08}\times
{\frac{\calL_{\rm J/\uppsi}}{\calL_{3.08}} \times
\frac{s_{3.08}}{s_{\rm J/\uppsi}}}=(118.66 \pm 0.05)\times 10^{6} ,
\end{eqnarray}
where $\calL_{\jpsi}$ and $\calL_{3.08}$ are the integrated
luminosities for the \jpsi data sample and the data sample
taken at $\sqrt{s}=3.08$~GeV, respectively, and $s_{\jpsi}$ and
$s_{3.08}$ are the corresponding squares of the center-of-mass
energies. The background is calculated to be $(1.717 \pm 0.002)\%$ of
the number of selected inclusive \jpsi events taken in
2017--2019.

\section{Determination of the detection efficiency and correction factor}
\label{eff-corr-factor}
In this analysis, the detection efficiency is determined
experimentally using a sample of \jpsi events from the
reaction $\rm \uppsi(3686) \rightarrow \uppi^+\uppi^- \jpsi$ to
reduce the uncertainty related to any discrepancies between the MC
simulation and the data. To ensure that the beam conditions and
detector status are similar to those of the sample collected at the
\jpsi peak, a dedicated $\rm \uppsi(3686)$ sample taken on
May 20, 2018 is used for this study.

For a candidate $\rm \uppsi(3686) \rightarrow \uppi^+\uppi^- \jpsi$
event, there must be at least two soft pions with opposite charge
detected in the MDC with $|\cos\theta| <
0.93$. Each candidate pion is required to have a momentum less than
$0.4\;\text{GeV}/c$, and the distance of closest approach to the IP
must satisfy $|V_z|<15$ cm and $V_r<1$ cm.  No further selection
criteria on the remaining charged tracks or showers are required. The
distribution of the invariant mass recoiling against all possible soft
$\rm \uppi^+\uppi^-$ pairs is shown in Fig.~\ref{fitdat}. A prominent
peak around $3.1\;\text{GeV}/c^2$, corresponding to the decay of $\rm
\uppsi(3686) \rightarrow \uppi^+\uppi^- \jpsi$ is observed over a
smooth background. The number of inclusive \jpsi events,
$N_\text{inc} = (3538.5 \pm 3.6)\times 10^{3}$, is obtained by fitting
a double-Gaussian function for the \jpsi signal plus a
second-order Chebychev polynomial for the background to the
$\uppi^+$$\uppi^-$ recoil mass spectrum.

To measure the detection efficiency of inclusive \jpsi
events, the same selection criteria as described in
Sec.~{\ref{hadsel}} are applied to the remaining charged tracks and
showers. The number of selected inclusive \jpsi
events, $N_\text{inc}^\text{sel}$, is determined to be
$(2717.6 \pm 3.4)\times 10^{3}$ using a fit to the recoil mass
distribution of the selected events with the same function as
described above.  The detection efficiency of inclusive \jpsi
events, $\epsilon^{\uppsi(3686)}_\text{data} = (76.80 \pm 0.05)\%$, is
calculated from the ratio of the number of inclusive \jpsi
events with and without the inclusive \jpsi event selection
criteria applied.

To account for the efficiency difference between the \jpsi
produced at rest and the \jpsi from the decay $\rm
\uppsi(3686) \rightarrow \uppi^+\uppi^- J/\uppsi$, a correction
factor, defined in Eq.~(\ref{Fcor}), is used. Two large statistics MC
samples, inclusive $\rm \uppsi(3686)$ and inclusive \jpsi
events, are produced and are subjected to the same selection criteria
as the data samples. The detection efficiencies of inclusive $\rm
J/\uppsi$ events are determined to be
$\epsilon^{\uppsi(3686)}_\text{MC} = (76.93 \pm 0.02)$\%, and
$\epsilon^{\rm J/\uppsi}_\text{MC}=(77.56 \pm 0.01 )$\% for the two
inclusive MC samples, respectively. The correction factor
$f_\text{cor}$ for the detection efficiency is therefore taken as

\begin{eqnarray}
\label{Fcorr} f_\text{cor} = \frac {\epsilon^{\rm J/\uppsi}_\text{MC}}
{\epsilon^{\uppsi(3686)}_\text{MC}}=1.0082 \pm 0.0007,
\end{eqnarray}
where the error is statistical only.
\begin{center}
\includegraphics[width=8.0cm,height=5.7cm]{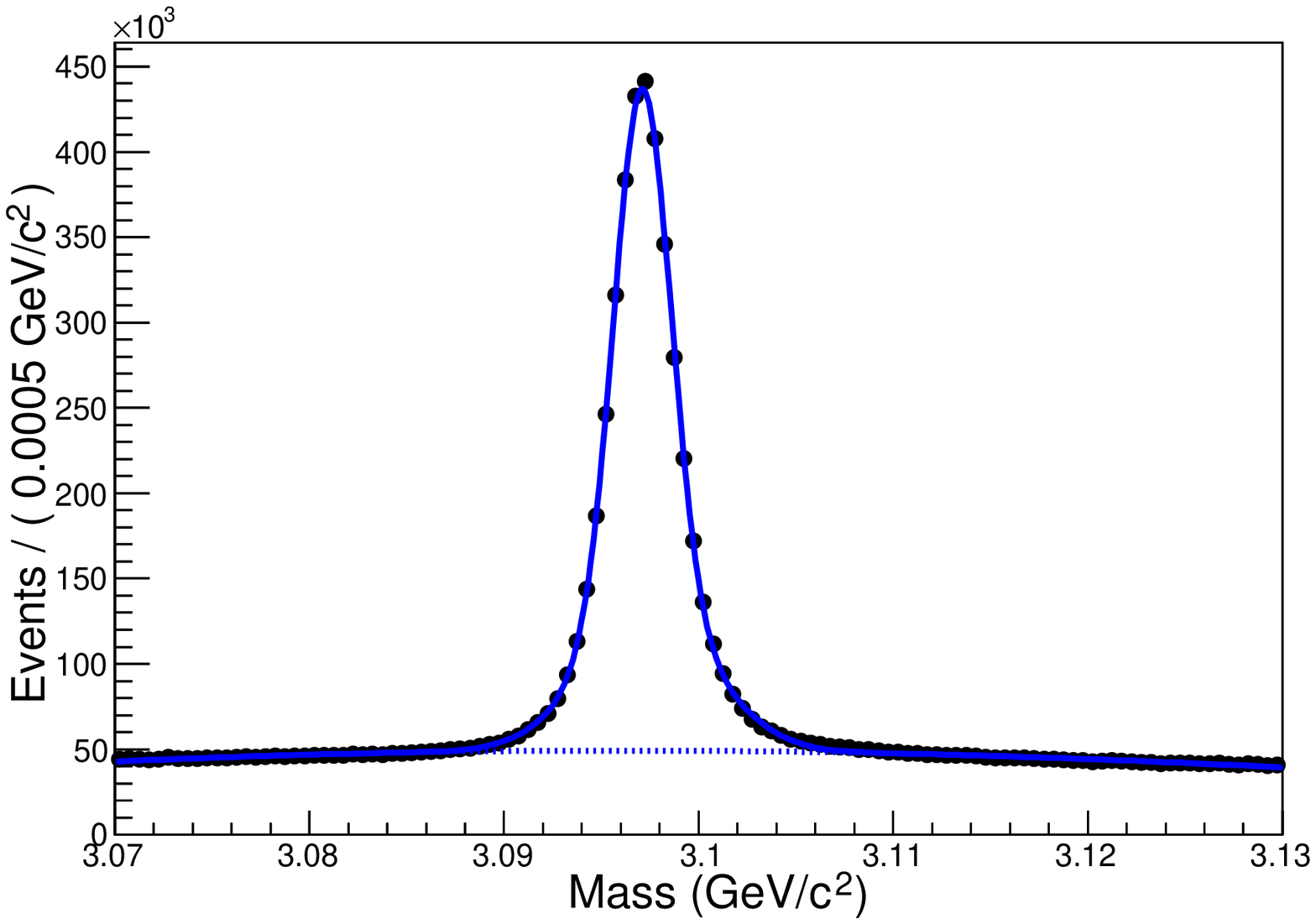}
\figcaption{\label{fitdat}Invariant mass recoiling against selected $\uppi^+\uppi^-$ pairs for the $\rm \uppsi(3686)$ data sample. The curves are the
results of the fit described in the text.}
\end{center}

\section{\boldmath Number of \jpsi events}

With Eq.~(\ref{Nojpsi}) and the corresponding parameter values
summarized in Table~\ref{Nformula}, the number of \jpsi
events collected in 2017--2019 is determined to be $(8774.0 \pm 0.2)
\times 10^6$. The trigger efficiency of the BESIII detector is taken
to be 100\%,
based on a study of various reactions~\cite{trig}. With the same
procedure, the numbers of \jpsi events taken in 2012 and 2009
are recalculated to be $(1088.5 \pm 0.1) \times 10^6$ and $(224.0 \pm
0.1) \times 10^6$, respectively, where the uncertainties are
statistical only. The statistical uncertainties of $N_\text{bg}$ are
taken into account as part of the systematic uncertainty (see
Sec.~{\ref{bkgerror}}). The systematic uncertainties from different
sources are discussed in detail in Sec.~{\ref{syst}}.
\begin{center}
\tabcaption{\label{Nformula}The values used in the calculation, and
  the resulting number of \jpsi events, where the
  uncertainties are statistical.}  \footnotesize
\begin{tabular*}{85mm}{l@{\extracolsep{\fill}}ccc}
\toprule Item &2017-2019 & 2012&2009\\ \hline
$N_\text{sel}(\times 10^6)$ & $6912.03 \pm 0.08$ &$860.59 \pm 0.03$ &$180.84 \pm 0.01$ \\
$N_\text{bg}(\times 10^6)$  & $118.66 \pm 0.05$ &$15.32 \pm 0.02$ &$6.89 \pm 0.04$ \\
$\epsilon_\text{trig}$ & 1.00  & 1.00 & 1.00 \\
$\epsilon^{\uppsi(3686)}_\text{data}$& $0.7680 \pm 0.0005$  & $0.7699 \pm 0.0005$ &$0.7707 \pm 0.0001$  \\
$\epsilon^{\uppsi(3686)}_\text{MC}$ & $0.7693 \pm 0.0002$ & $0.7709 \pm 0.0002$ & $0.7723 \pm 0.0002$ \\
$\epsilon^{\rm J/\uppsi}_\text{MC}$ & $0.7756 \pm 0.0001$ &  $0.7776 \pm 0.0001$ &$0.7780 \pm 0.0001  $ \\
$f_\text{cor}$ &  $1.0082 \pm 0.0007$  &  $1.0086 \pm 0.0008$ &$1.0074 \pm 0.0003$ \\\midrule
$N_{\rm J/\uppsi}(\times 10^6)$ & $8774.0 \pm 0.2$&$1088.5 \pm 0.1$
&$224.0 \pm 0.1$  \\
\bottomrule
\end{tabular*}
\end{center}

\section{Systematic uncertainty}
\label{syst}

The sources of systematic uncertainties, including the MC model, track
reconstruction efficiency, fit to the \jpsi peak, background
estimation, random trigger mixing and the efficiency of selecting the two soft
pions recoiling against \jpsi, are investigated in detail
below, and the corresponding contributions are summarized in
Table~\ref{TABSYS}.

\subsection {MC model uncertainty}

In the measurement of the number of \jpsi events, only the
efficiency correction factor, $f_\text{cor}$, depends on the MC
simulation.  To evaluate the uncertainty due to the MC model, MC
samples are generated with the incomplete MC model, and the correction factor
based on these samples is compared to its nominal value. As shown in
Fig.~\ref{ngood}, the charged track multiplicity distribution of the
incomplete MC sample deviates much from the experimental data, which means
this method will overestimate the systematic uncertainty.  To be
conservative, the change in the correction factor, 0.18\%, is taken as
the systematic uncertainty due to the MC model on the number of
\jpsi events collected in 2017--2019 (0.18\% for 2012 and 0.27\%
for 2009).

\subsection {Track reconstruction efficiency}

The charged track reconstruction efficiencies in MC simulation and
experimental data are studied, and the disagreement between them is
less than $1$\% for each charged track. In the analysis, the detection
efficiency for inclusive \jpsi decays is obtained using the
$\rm \uppsi(3686)$ data sample. The consistency of charged track
reconstruction efficiency between the MC and data samples in $\rm
\uppsi(3686)$ decays is assumed to be the same as that in
\jpsi decays since the $\rm \uppsi(3686)$ data is taken in close
chronological proximity to the \jpsi sample. To evaluate the
effect of a possible difference, the track reconstruction efficiencies
in both \jpsi and $\rm \uppsi(3686)$ MC samples are varied by
$-1\%$ to determine the uncertainty due to the MDC tracking.  As
expected, the change in the correction factor is very small, $0.02\%$,
and this value is taken as a systematic uncertainty ($0.03\%$ for
2012).

In 2009, the \jpsi and $\rm \uppsi(3686)$ data samples were
collected in different time periods, and there may be slight
differences in the tracking efficiencies between the two data sets.
Here, the difference between the MC/data consistencies in the
\jpsi and $\rm \uppsi(3686)$ samples is assumed to be 0.5\%, half
of the data/MC inconsistency, 1\%.  To estimate the corresponding
systematic uncertainty, we modify the track reconstruction efficiency
by $-0.5\%$ in the \jpsi MC sample, keeping it unchanged for
the $\rm \uppsi(3686)$ MC sample. The resulting change in the
correction factor, $0.31\%$, is taken as a systematic uncertainty on
the number of \jpsi events in 2009.

\subsection{\boldmath Fit to the \jpsi peak}

The $\rm \uppsi(3686)$ data sample is used to measure the selection
efficiency of inclusive \jpsi events. The yield of \jpsi
events in $\rm \uppsi(3686)$ decays is determined by fitting
the \jpsi peak in the mass spectrum recoiling against $\uppi^+\uppi^-$. The
uncertainties due to the fit are investigated: $(a)$ \emph{the~fit}:
we propagate the statistical uncertainties of the \jpsi
signal yield from the fit to the selection efficiency, and the
resulting uncertainties, 0.07\% and 0.03\% for
$\epsilon^{\uppsi(3686)}_{data}$ and $\epsilon^{\uppsi(3686)}_{MC}$,
respectively, are considered to be the uncertainties from the fit
itself. $(b)$ \emph{the~fit~range}: the fit range on the
$\uppi^+\uppi^-$ recoil mass is changed from [3.07, 3.13] GeV/$c^2$
to [3.08, 3.12] GeV/$c^2$, and the change of the result, 0.07\%, is
taken as the corresponding systematic uncertainty. $(c)$
\emph{the~signal~description}: we perform an alternative fit by
describing the \jpsi signal with a histogram (convolved with
a Gaussian function) obtained from the recoil mass spectrum of $\rm
\uppi^+ \uppi^-$ in $\rm \uppsi(3686)\rightarrow \uppi^+ \uppi^-
\jpsi, ~\jpsi \rightarrow \upmu^+\upmu^-$, and the resulting
change, 0.01\%, is considered to be the associated systematic
uncertainty. $(d)$ \emph{the~background~shape}: the uncertainty due to
the background shape, 0.03\%, is estimated by replacing the
second-order Chebychev polynomial with a first-order or third-order
Chebychev polynomial. By assuming that all of the sources of
systematic uncertainty are independent, the fit uncertainty for the
2017--2019 \jpsi sample, 0.10\%, is obtained by adding all of
the above effects in quadrature.

The same sources of systematic uncertainty are considered for the
\jpsi sample taken in 2012 (2009).  The fit has an uncertainty of
0.07\% (0.02\%) for $\epsilon^{\uppsi(3686)}_\text{data}$ and 0.03\%
(0.03\%) for $\epsilon^{\uppsi(3686)}_\text{MC}$. The uncertainties
from the fit range, signal function and background shape are 0.08\%
(0.03\%), 0.15\% (0.06\%) and 0.10\% (0.04\%), respectively. The total
uncertainty from the fit for the 2012 (2009) data is 0.21\% (0.09\%).

\subsection{Background uncertainty}
\label{bkgerror}
In this analysis, the events selected from the experimental data
sample include the \jpsi events and background: QED
processes, cosmic rays, beam-induced backgrounds, and electronic noise. The
contribution of the background is estimated by normalizing the number
of events in the continuum data sample taken at $\sqrt{s} = 3.08$ GeV
according to Eq.~(\ref{Nbg}). The uncertainty due to the background
estimation mainly comes from the normalization method, the statistics
of the continuum sample, the statistical uncertainty of the integrated
luminosity and the uncertainty due to beam associated backgrounds.

The cosmic ray background, beam associated backgrounds and
electronic noise can not be normalized properly with Eq.~(\ref{Nbg}), since
the number of cosmic rays is proportional to the time of data taking,
while beam associated backgrounds depend on the vacuum status and beam
currents during data taking in addition to the time of data taking, and
the electronic noise also depends on the detector status.  To estimate
the associated systematic uncertainty, the difference in the estimated
number of background events with and without the
energy-dependent factor in Eq.~(\ref{Nbg}) is used.

During 2017--2019, two data samples at $\sqrt{s} = 3.08$ GeV were
taken at different times during the \jpsi data taking. They are
compared to each other to estimate the uncertainty of the background
related with the stability of the beam and vacuum status. Each of the
two continuum data samples is used to estimate the background with
Eq.~(\ref{Nbg}), and the maximum difference to the nominal result,
0.03\%, is taken as the related systematic uncertainty. The
corresponding systematic uncertainty for the 2012 sample is 0.09\%. Only
one continuum data sample was taken for \jpsi data in
2009. The selected background events from the continuum sample are
compared to those from the \jpsi data to estimate the
corresponding uncertainty as described in detail in
Ref. \cite{njsi2009}.

Assuming that all the above effects are independent, their
contributions are added quadratically. The resulting uncertainties on
the number of \jpsi events due to the background estimation
are determined to be 0.04\%, 0.10\% and 0.14\% for the data taken in
2017--2019, 2012 and 2009, respectively.

\subsection {Random Trigger mixing}

In the MC simulation, events recorded by a random trigger are mixed
into the MC events to simulate the electronic noise and beam-induced
background. In the $\rm \uppsi(3686)$ MC sample the random trigger
events from the $\rm \uppsi(3686)$ data taking were replaced by the
random trigger events from the \jpsi data taking to estimate
the effect of the different background levels. The change of the correction
factor for the detection efficiency, 0.06\%, is taken as the
systematic uncertainty due to random trigger mixing for the number of
\jpsi events taken in 2017--2019. The corresponding
uncertainties for the 2012 and 2009 samples are 0.02\% and 0.12\%,
respectively.

\subsection {\boldmath Uncertainty of selection efficiency of two soft pions}
 Study of the MC sample shows that the selection efficiency of soft
 pions, $\epsilon_{\uppi^+\uppi^-}$, recoiling against the $\rm
 J/\uppsi$ in $\rm \uppsi(3686)\rightarrow \uppi^+\uppi^-J/\uppsi$
 depends on the multiplicity of charged tracks in the \jpsi
 decays.  Differences between the data and MC samples may lead to a
 change in the number of \jpsi events.  The dependence of
 $\epsilon_{\uppi^+\uppi^-}$ in the data is obtained by comparing the
 multiplicity distribution of \jpsi decays in the $\rm
 \uppsi(3686) \rightarrow \uppi^+\uppi^- J/\uppsi$ data sample to that
 of the \jpsi data at rest.  Then the efficiency of $\rm
 J/\uppsi$ in the $\rm \uppsi(3686) \rightarrow \uppi^+ \uppi^-
 J/\uppsi(J/\uppsi\rightarrow \text{inclusive})$ MC sample,
 $\epsilon^{\uppsi(3686)}_\text{MC}$ in Eq.~(\ref{Fcor}), can be
 reweighted with the dependence of $\epsilon_{\uppi^+\uppi^-}$ from
 the data sample. The resulting changes in the number of $\rm
 J/\uppsi$ events, 0.40\%, 0.27\%, 0.32\% are taken as the
 uncertainties for the data taken in 2017--2019, 2012 and 2009,
 respectively.

\subsection{Summary of systematic uncertainties}
The systematic uncertainties from the different sources studied above
are summarized in Table~\ref{TABSYS}. The total systematic uncertainty
for the number of \jpsi events in 2017--2019, 0.45\%, is the
quadratic sum of the individual uncertainties.  Correspondingly, the
uncertainties for 2012 and 2009 are 0.40\% and 0.56\%, respectively.

\begin{center}
  \tabcaption{\label{TABSYS}Sources of systematic uncertainties
    and the corresponding
  contributions to the number of \jpsi events, where the
  superscript * means the error is common for the same item in
  different data samples.}  \footnotesize
\begin{tabular*}{82mm}{l@{\extracolsep{\fill}}ccc}
\toprule Sources & 2017--2019(\%) &2012 (\%) & 2009(\%)\\ \hline
MC model uncertainty & 0.18$^*$ &0.18$^*$ &0.27$^*$ \\
Tracking efficiency &0.02$^*$& 0.03$^*$&0.31 \\
Fit to \jpsi peak &0.10  & 0.21&0.09 \\
Background uncertainty & 0.04 & 0.10 &0.14 \\
Noise mixing & 0.06 & 0.02& 0.12 \\
$\epsilon_{\uppi^+\uppi ^-}$ uncertainty  &0.40$^*$ &0.27$^*$& 0.32$^*$ \\\hline
Total &0.45& 0.40&0.56 \\
\bottomrule
\end{tabular*}
\end{center}

\section{Summary}
Using the inclusive \jpsi decays, the number of \jpsi
events collected with the BESIII detector in 2017--2019 is
determined to be $ (8774.0 \pm 39.4)\times10^{6},$ where the
uncertainty is completely dominated by systematics and the statistical
uncertainty is negligible.  The numbers of \jpsi events taken
in 2009 and 2012 are recalculated to be $(224.0 \pm 1.3)\times10^{6}$
and $ (1088.5 \pm 4.4)\times10^{6},$ which are consistent with the
previous measurements~\cite{njsi2012}, but with improved precision.

The total number of \jpsi events taken with BESIII
detector is determined to be $ N_{\rm J/\uppsi}= (10087 \pm
44)\times10^{6}$.  Here, the total uncertainty is determined by adding
the common uncertainties linearly and the independent ones in
quadrature.

\vspace{6mm}
{\it  The BESIII collaboration thanks the staff of BEPCII and the IHEP computing center for their strong support.}

\end{multicols}

\clearpage

\vspace{6mm}


\begin{thebibliography}{90}

\vspace{3mm}

\bibitem{white}A review of the recent results can be found in the bibliography of Chapter 2 of: M. Ablikim et al. [BESIII Collaboration], Chin. Phys. C {\bf 44}, 040001 (2020)
\bibitem{bes3}M. Ablikim et al (BESIII Collaboration), Nucl. Instrum. Methods A, {\bf 614}: 345-399 (2010)
\bibitem{njsi2012}M. Ablikim et al (BESIII Collaboration), Chin. Phys. C, {\bf 41}(1): 013001 (2017)
\bibitem{kkmc} S. Jadach, B. F. L. Ward, Z. Was, Comput. Phys. Commu. {\bf
130}:130 (2000); S. Jadach, B. F. L. Ward, Z. Was, Phys. Rev. D, {\bf 63}: 113009 (2001)
\bibitem{evtgen}R. G. Ping, HEP \& NP, {\bf 32}(8): 599-602 (2008)
\bibitem{djl}D. J. Lange, Nucl. Instrum. Methods A, {\bf 462}: 152-155 (2001)
\bibitem{PDG} P.A. Zyla et al. (Particle Data Group), Prog. Theor. Exp. Phys. 2020, 083C01 (2020)
\bibitem{LUND}J. C. Chen et al, Phys. Rev. D, {\bf 62}: 034003 (2000)
\bibitem{LUND2} R. L. Yang, R. G. Ping, H. Chen, Chin. Phys. Lett., {\bf
31}: 061301 (2014)
\bibitem{geant4} S. Agostinelli et al,  Nucl. Instrum. Methods A, {\bf
506}: 250-303 (2003)
\bibitem{trig}  M. Ablikim  et al.  (BESIII Collaboration), Chin. Phys. C {\bf 45}, 023002 (2021); N. Berger et al, Chin. Phys. C, {\bf 34}(12): 1779-1784 (2010)
\bibitem{njsi2009}M. Ablikim et al (BESIII Collaboration), Chin. Phys. C, {\bf 36}(10): 915-925 (2012)




\end{thebibliography}
\end{document}